\documentclass{moriond}

\def\be{\begin{equation}}
\def\ee{\end{equation}}
\def\bea{\begin{eqnarray}}
\def\eea{\end{eqnarray}}

\usepackage{lineno}
\usepackage{xspace}
\usepackage{xcolor}
\usepackage{hyperref}
\usepackage{subfigure}
\usepackage{graphicx}
\usepackage{lineno}

\usepackage{listings}
\usepackage{amsmath}
\usepackage{afterpage}
\usepackage[colorlinks=true]{hyperref}
\usepackage{makeidx}

\newcommand{\jpsi} {\ensuremath{{\mathrm J}/\psi}\xspace}
\newcommand{\psip} {\ensuremath{\psi'}\xspace}

\newcommand{\twohundrednn}       {$\sqrt{s_{\mathrm{NN}}}~=~200$~Ge\kern-.1emV\xspace}

\newcommand{\starlight}{STARlight\xspace}

\begin{document}
%

\newcommand{\pp}           {pp\xspace}
\newcommand{\ppbar}        {\mbox{$\mathrm {p\overline{p}}$}\xspace}
\newcommand{\XeXe}         {\mbox{Xe--Xe}\xspace}
\newcommand{\PbPb}         {\mbox{Pb--Pb}\xspace}
\newcommand{\pA}           {\mbox{pA}\xspace}
\newcommand{\pPb}          {\mbox{p--Pb}\xspace}
\newcommand{\AuAu}         {\mbox{Au--Au}\xspace}
\newcommand{\dAu}          {\mbox{d--Au}\xspace}

\newcommand{\s}            {\ensuremath{\sqrt{s}}\xspace}
\newcommand{\snn}          {\ensuremath{\sqrt{s_{\mathrm{NN}}}}\xspace}
\newcommand{\pt}           {\ensuremath{p_{\rm T}}\xspace}
\newcommand{\meanpt}       {$\langle p_{\mathrm{T}}\rangle$\xspace}
\newcommand{\ycms}         {\ensuremath{y_{\rm CMS}}\xspace}
\newcommand{\ylab}         {\ensuremath{y_{\rm lab}}\xspace}
\newcommand{\y}         {\ensuremath{y}\xspace}
\newcommand{\etarange}[1]  {\mbox{$\left | \eta \right |~<~#1$}}
\newcommand{\yrange}[1]    {\mbox{$\left | y \right |~<~#1$}}
\newcommand{\dndy}         {\ensuremath{\mathrm{d}N_\mathrm{ch}/\mathrm{d}y}\xspace}
\newcommand{\dndeta}       {\ensuremath{\mathrm{d}N_\mathrm{ch}/\mathrm{d}\eta}\xspace}
\newcommand{\avdndeta}     {\ensuremath{\langle\dndeta\rangle}\xspace}
\newcommand{\dNdy}         {\ensuremath{\mathrm{d}N_\mathrm{ch}/\mathrm{d}y}\xspace}
\newcommand{\Npart}        {\ensuremath{N_\mathrm{part}}\xspace}
\newcommand{\Ncoll}        {\ensuremath{N_\mathrm{coll}}\xspace}
\newcommand{\dEdx}         {\ensuremath{\textrm{d}E/\textrm{d}x}\xspace}
\newcommand{\RpPb}         {\ensuremath{R_{\rm pPb}}\xspace}

\newcommand{\nineH}        {$\sqrt{s}~=~0.9$~Te\kern-.1emV\xspace}
\newcommand{\seven}        {$\sqrt{s}~=~7$~Te\kern-.1emV\xspace}
\newcommand{\eight}        {$\sqrt{s}~=~8$~Te\kern-.1emV\xspace}
\newcommand{\twoH}         {$\sqrt{s}~=~0.2$~Te\kern-.1emV\xspace}
\newcommand{\twosevensix}  {$\sqrt{s}~=~2.76$~Te\kern-.1emV\xspace}
\newcommand{\five}         {$\sqrt{s}~=~5.02$~Te\kern-.1emV\xspace}
\newcommand{\fiveExactly}  {$\sqrt{s}~=~5$~Te\kern-.1emV\xspace}
\newcommand{\twosevensixnn}{$\sqrt{s_{\mathrm{NN}}}~=~2.76$~Te\kern-.1emV\xspace}
\newcommand{\fivenn}       {$\sqrt{s_{\mathrm{NN}}}~=~5.02$~Te\kern-.1emV\xspace}
\newcommand{\LT}           {L{\'e}vy-Tsallis\xspace}
\newcommand{\GeVc}         {Ge\kern-.1emV/$c$\xspace}
\newcommand{\MeVc}         {Me\kern-.1emV/$c$\xspace}
\newcommand{\TeV}          {Te\kern-.1emV\xspace}
\newcommand{\GeV}          {Ge\kern-.1emV\xspace}
\newcommand{\GeVtwo}       {Ge\kern-.1emV$^2$\xspace}
\newcommand{\MeV}          {Me\kern-.1emV\xspace}
\newcommand{\GeVmass}      {Ge\kern-.2emV/$c^2$\xspace}
\newcommand{\MeVmass}      {Me\kern-.2emV/$c^2$\xspace}
\newcommand{\lumi}         {\ensuremath{\mathcal{L}}\xspace}

\newcommand{\ITS}          {\rm{ITS}\xspace}
\newcommand{\TOF}          {\rm{TOF}\xspace}
\newcommand{\ZDC}          {\rm{ZDC}\xspace}
\newcommand{\ZDCs}         {\rm{ZDCs}\xspace}
\newcommand{\ZNA}          {\rm{ZNA}\xspace}
\newcommand{\ZNC}          {\rm{ZNC}\xspace}
\newcommand{\SPD}          {\rm{SPD}\xspace}
\newcommand{\SDD}          {\rm{SDD}\xspace}
\newcommand{\SSD}          {\rm{SSD}\xspace}
\newcommand{\TPC}          {\rm{TPC}\xspace}
\newcommand{\TRD}          {\rm{TRD}\xspace}
\newcommand{\VZERO}        {\rm{V0}\xspace}
\newcommand{\VZEROA}       {\rm{V0A}\xspace}
\newcommand{\VZEROC}       {\rm{V0C}\xspace}
\newcommand{\Vdecay} 	   {\ensuremath{V^{0}}\xspace}

\newcommand{\pip}          {\ensuremath{\pi^{+}}\xspace}
\newcommand{\pim}          {\ensuremath{\pi^{-}}\xspace}
\newcommand{\kap}          {\ensuremath{\rm{K}^{+}}\xspace}
\newcommand{\kam}          {\ensuremath{\rm{K}^{-}}\xspace}
\newcommand{\pbar}         {\ensuremath{\rm\overline{p}}\xspace}
\newcommand{\kzero}        {\ensuremath{{\rm K}^{0}_{\rm{S}}}\xspace}
\newcommand{\lmb}          {\ensuremath{\Lambda}\xspace}
\newcommand{\almb}         {\ensuremath{\overline{\Lambda}}\xspace}
\newcommand{\Om}           {\ensuremath{\Omega^-}\xspace}
\newcommand{\Mo}           {\ensuremath{\overline{\Omega}^+}\xspace}
\newcommand{\X}            {\ensuremath{\Xi^-}\xspace}
\newcommand{\Ix}           {\ensuremath{\overline{\Xi}^+}\xspace}
\newcommand{\Xis}          {\ensuremath{\Xi^{\pm}}\xspace}
\newcommand{\Oms}          {\ensuremath{\Omega^{\pm}}\xspace}
\newcommand{\degree}       {\ensuremath{^{\rm o}}\xspace}
\vspace*{4cm}
\title{Recent results on \jpsi photoproduction in ultra-peripheral collisions with ALICE}

\author{ S. Ragoni, on behalf of the ALICE Collaboration }

\address{Creighton University, 2500 California Plaza,\\
Omaha, NE 68178, United States}

\maketitle\abstracts{
Ultra-peripheral collisions (UPC) are events characterised by large impact parameters between the two projectiles, larger than the sum of their radii. In UPCs, the protons and ions accelerated by the LHC do not interact via the strong interaction and can be regarded as  sources of quasireal photons. 
Photoproduction of vector mesons (e.g. \jpsi and \psip) in UPC is quite interesting since it is sensitive to the low-$x$ gluon density. 
Owing to the statistics available from Run 2 data, the ALICE Collaboration has been able to carry out differential measurements, such as the measurement of the rapidity-differential cross section of  coherent \jpsi production in Pb--Pb UPCs and the $t$-dependence of coherent and incoherent \jpsi, which are sensitive to the average of the spatial distribution of the gluons and its variance, respectively. 
In addition, the ALICE Collaboration has measured the energy dependence of the photonuclear cross section with a few different strategies, such as by measuring the rapidity-differential cross section of coherent \jpsi production in different impact parameter ranges i.e. in ultra-peripheral and peripheral collisions, and in neutron emission classes. The latter two provide promising tools for the resolution of the ambiguity in Bjorken-$x$ which arises in symmetric A--A UPCs.  
}

\section{Introduction}
Vector meson photoproduction in \PbPb ultra-peripheral collisions (UPCs) is currently being actively studied at the LHC. In this type of events the two nuclei lie at impact parameters larger than the sum of their radii. A photon from one nucleus interacts with a colourless object from the other nucleus, a Pomeron, and a vector meson is likely the result of such an interaction. The large data set collected during LHC Run 2 enabled the ALICE Collaboration to perform several differential measurements. It has been possible to study \jpsi photoproduction in terms of $\pt^{2}$ - the vector meson transverse momentum squared - and in terms of its rapidity, $y$. More recently, coherent \jpsi production  allowed to measure the energy dependence of the photonuclear cross section and of the nuclear suppression factor down to a Bjorken-$x$ of about $10^{-5}$. 

\section{Studying coherent and incoherent \jpsi production as a function of $\pt^{2}$}
\begin{figure}
	\begin{center}
		\subfigure[]{
			\label{fig:coh-pt}
			\includegraphics[width=0.45\textwidth]{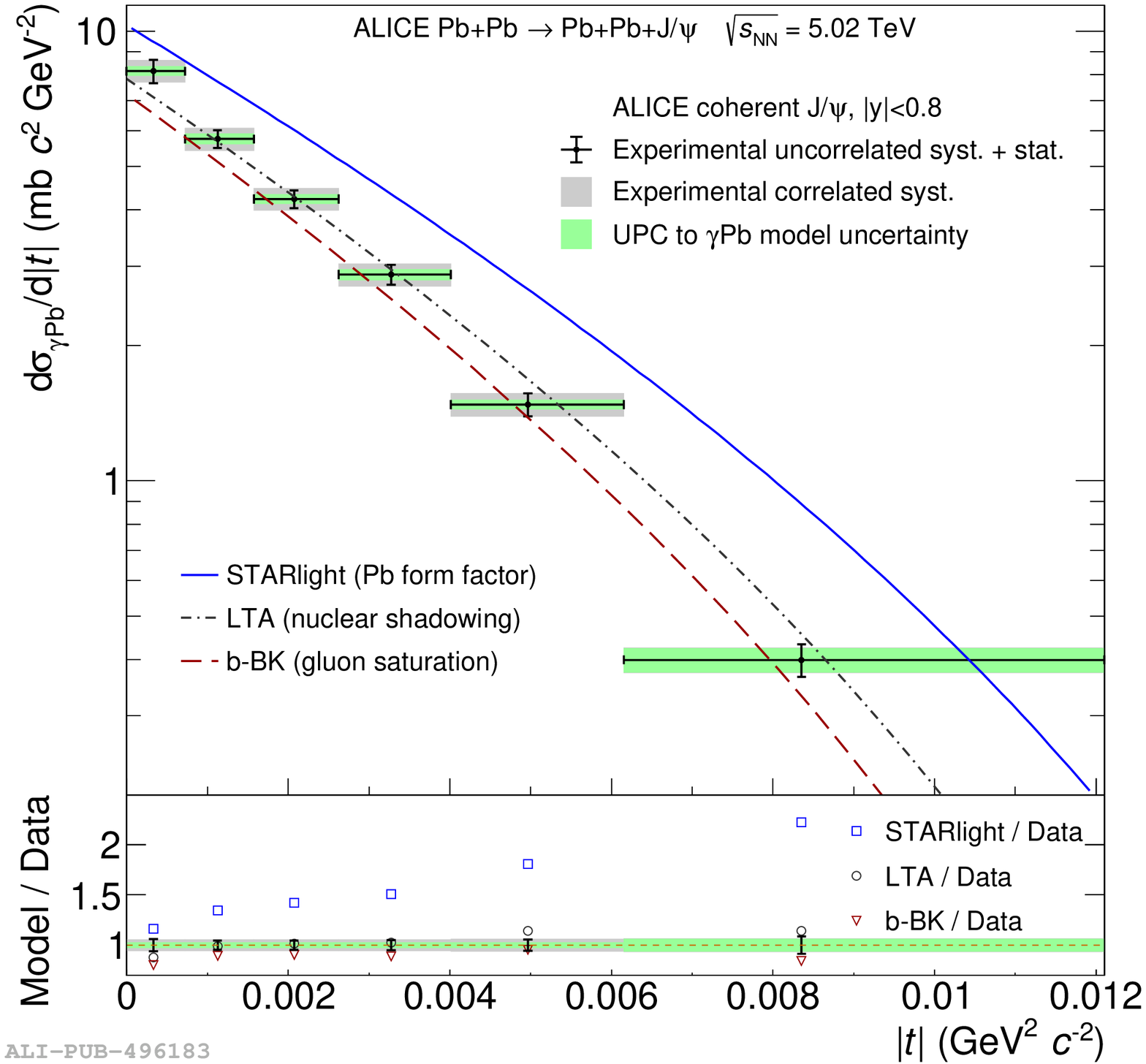}
		}
		\subfigure[]{
			\label{fig:incoh-pt}
			\includegraphics[width=0.45\textwidth]{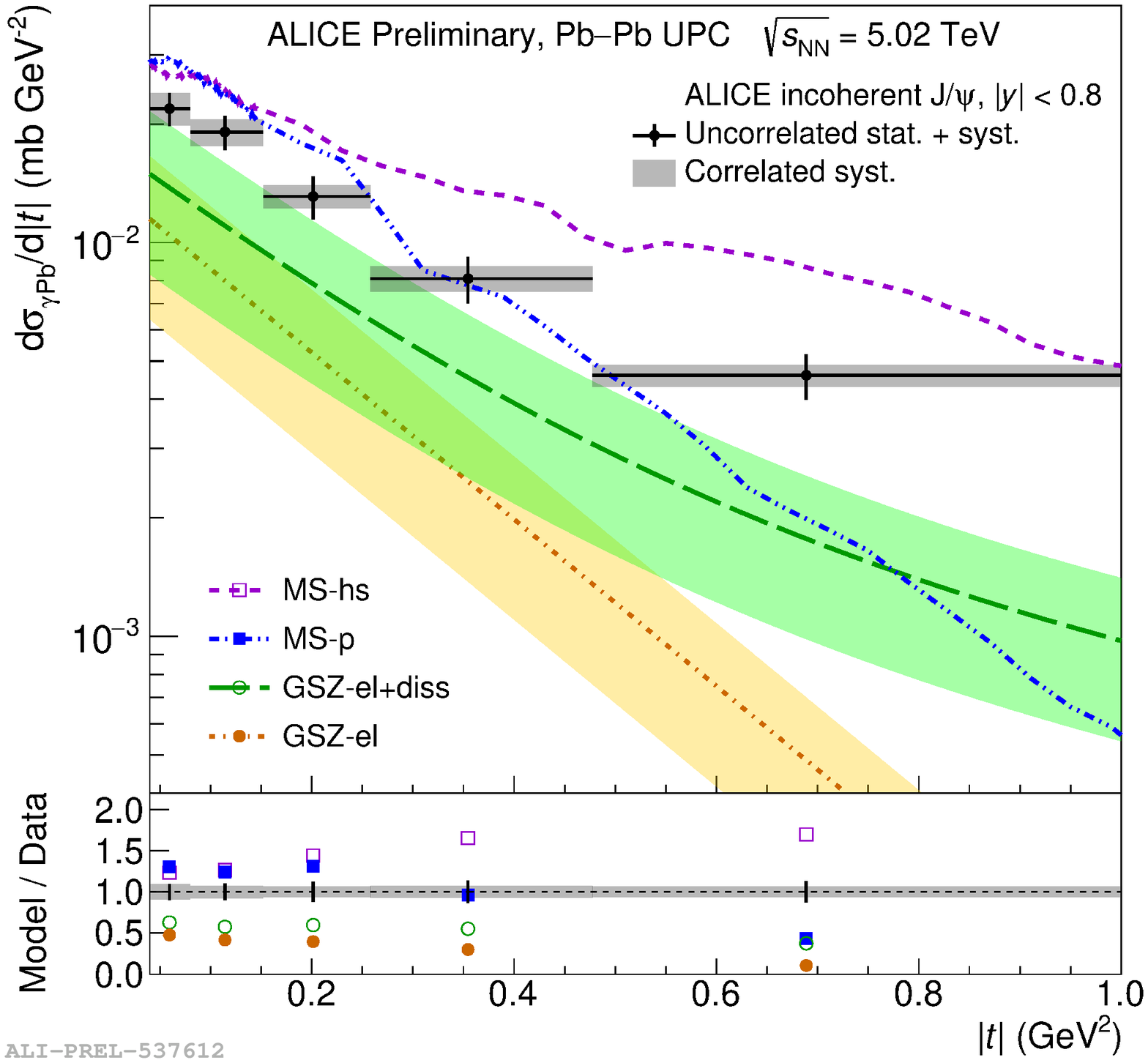}
		}\\ 
	\end{center}
	\caption{$|t|$-dependence of coherent and incoherent \jpsi photonuclear production in Fig.~\ref{fig:coh-pt} and \ref{fig:incoh-pt}, respectively. From the former it is evident how models involving QCD dynamical effects are needed to describe the data, while from the latter it can be observed that models containing subnucleonic degrees of freedom better describe the slope of the data.  }
	\label{fig:pt}
\end{figure}

Using the data from LHC Run 2 it was possible to measure (in)coherent \jpsi production as a function of $\pt^{2} \approx |t|$, where $|t|$ is the square of the momentum transferred to the target nucleus. 
This observable has remarkable applications: the $|t|$-dependence of coherent \jpsi is sensitive to the average of the spatial distribution of the gluons~\cite{Good:1960ba}, while that of incoherent \jpsi is sensitive to its variance instead~\cite{Miettinen:1978jb}. Fig.~\ref{fig:coh-pt} shows the results for coherent \jpsi measured with the ALICE detector~\cite{ALICE:2021tyx}. The data are described by models including QCD dynamical effects. Fig.~\ref{fig:incoh-pt} shows the first measurement of the $|t|$-dependence of incoherent \jpsi. While none of the models manages to describe both the slope and the normalisation of the data of the distribution for incoherent \jpsi, it can be observed that only those models including gluonic subnucleon fluctuations manage to reproduce the slope of the data (such as the MS-hs and GSZ-el+diss models). It is interesting to note that the measurement of the $|t|$-dependence of incoherent \jpsi production will also be a powerful observable to measure gluon saturation~\cite{Cepila:2016uku} in the future with LHC Run 3 data.

\section{Studying coherent \jpsi production as a function of $y$}
\begin{figure}
    \centering \includegraphics[width=0.45\textwidth]{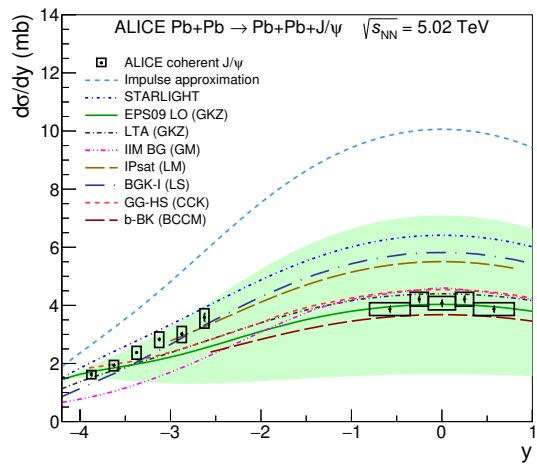}
    \caption{$y$-dependence of coherent \jpsi production in \PbPb ultra-peripheral collisions. The data show important shadowing effects. }
    \label{fig:coh-y}
\end{figure}

The rapidity dependence of coherent \jpsi production was measured at forward~\cite{ALICE:2019tqa} and midrapidity~\cite{ALICE:2021gpt}. The ALICE data are compared with models in Fig.~\ref{fig:coh-y}, and the data favour those models featuring moderate nuclear shadowing.

\section{Studying coherent \jpsi production as a function of $W_{\gamma {\rm Pb}, n}^2$}
\begin{figure}
	\begin{center}
		\subfigure[]{
			\label{fig:xsec-energy}
			\includegraphics[width=0.45\textwidth]{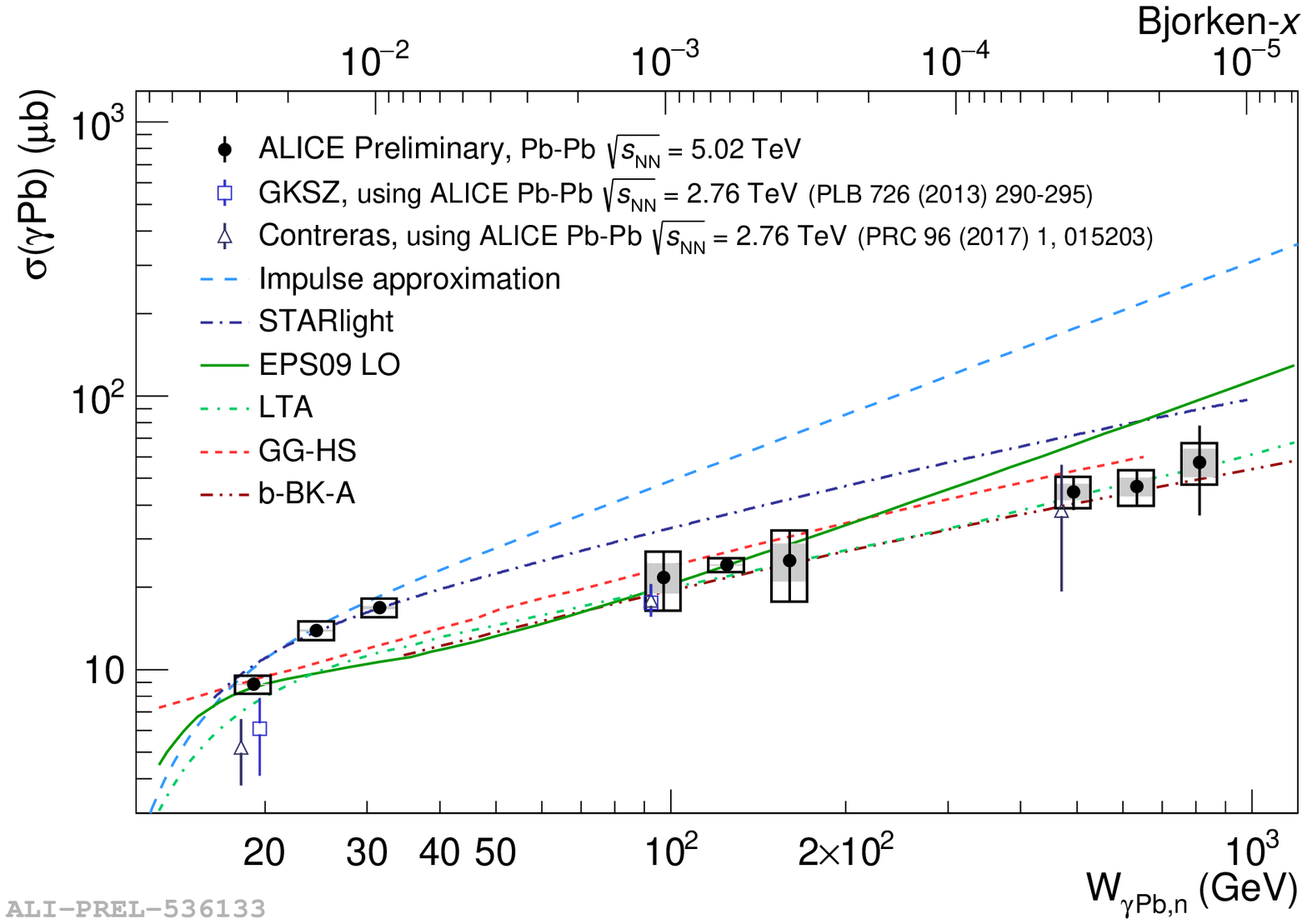}
		}
		\subfigure[]{
			\label{fig:shadowing}
			\includegraphics[width=0.45\textwidth]{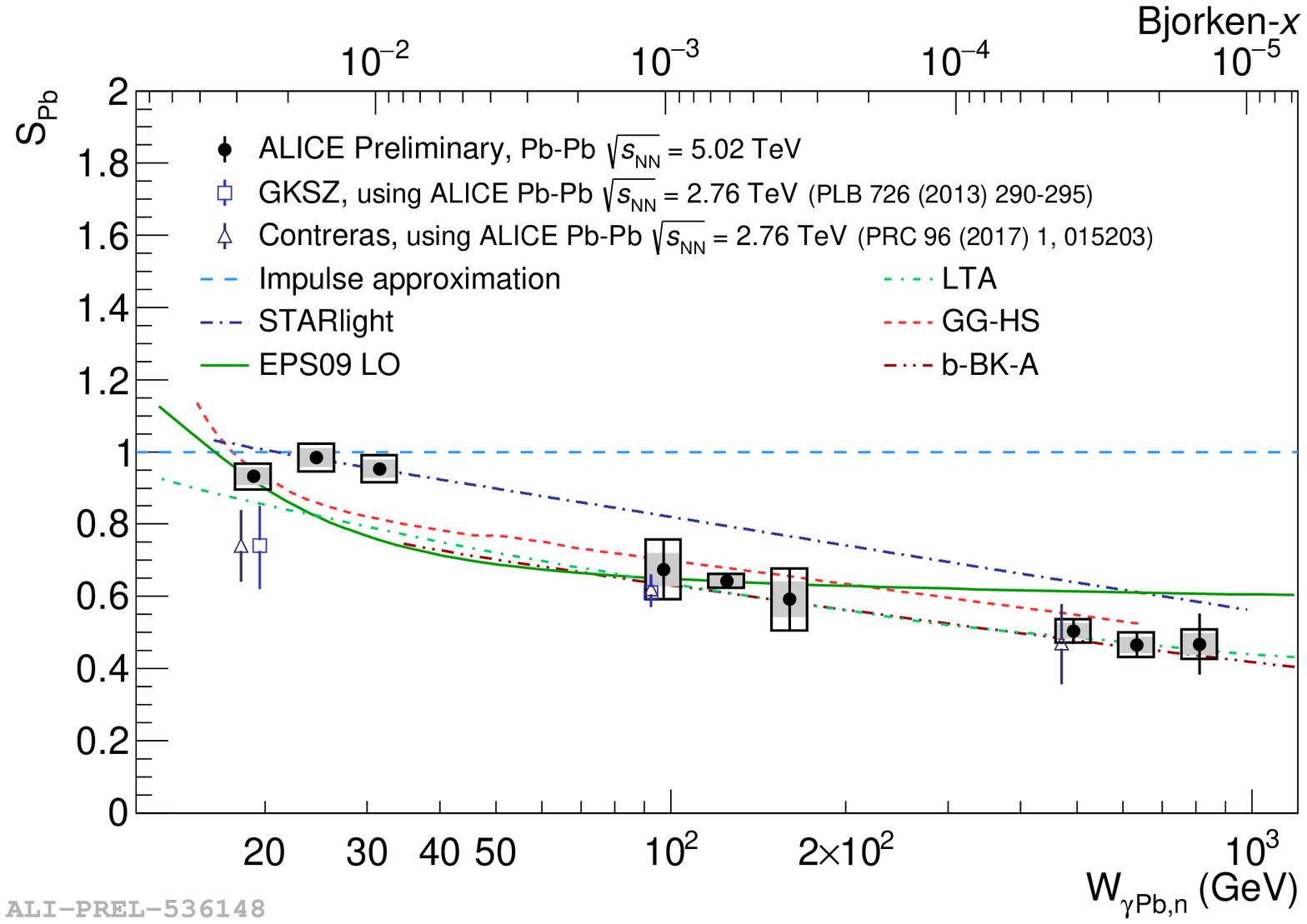}
		}\\ 
	\end{center}
	\caption{The energy dependence of the photonuclear cross section and of the nuclear suppression factor are shown in Fig.~\ref{fig:xsec-energy} and \ref{fig:shadowing}, respectively. ${\rm W}_{\gamma {\rm Pb,n}}$ is the photon--nucleon center of mass energy.  }
	\label{fig:energy-dependence}
\end{figure}

The production cross section for coherent \jpsi in \PbPb UPCs is described by Eq.~\ref{eq:photoproduction}:
\begin{equation}
	\label{eq:photoproduction}
		\frac{{\rm d}\sigma_{{\rm Pb}{\rm Pb}}}{{\rm d}y} = n(\gamma, +y)\cdot { \sigma_{\gamma {\rm Pb}}(+y)} + n(\gamma, -y)\cdot {\sigma_{\gamma {\rm Pb}}(-y)} \text{ ,} 
\end{equation}
where ${\rm d}\sigma_{{\rm Pb}{\rm Pb}}/{\rm d}y$ is the production cross section for coherent \jpsi, $n(\gamma, \pm y)$ are the photon fluxes and $\sigma_{\gamma {\rm Pb}}(\pm y)$ are the photonuclear cross sections. Thus, at midrapidity i.e. $y\sim 0$ there is no ambiguity and the coherent cross section is directly related to the photonuclear cross section. At forward rapidity the contribution from the high-$x$ photon flux is much larger than the low-$x$ solution~\cite{Baltz:2002pp}, so that the photonuclear cross section at high-$x$ can be approximated by the coherent cross section modulo the photon flux. To resolve the low-$x$ solution, i.e. to be able to reach down to $x\sim 10^{-5}$ with the pseudorapidity acceptance of the ALICE detector, it is possible to separate the dataset in neutron emission classes and measure coherent \jpsi production as a function of neutron emission, as shown in Eq.~\ref{eq:neutron-emission-splitting}~\cite{Guzey:2013jaa}:
\begin{equation}
	\label{eq:neutron-emission-splitting}
	\begin{split}
		\frac{{\rm d}\sigma^{0{\rm N}0{\rm N}}_{{\rm Pb}{\rm Pb}}}{{\rm d}y} &= n_{0{\rm N}0{\rm N}}(\gamma, +y) \cdot {\sigma_{\gamma {\rm Pb}}(+y)} +  n_{0{\rm N}0{\rm N}}(\gamma, -y) \cdot {\sigma_{\gamma {\rm Pb}}(-y)}  \text{ ,}\\
		\frac{{\rm d}\sigma^{0{\rm NX}{\rm N}}_{{\rm Pb}{\rm Pb}}}{{\rm d}y}  &= n_{0{\rm NX}{\rm N}}(\gamma, +y) \cdot {\sigma_{\gamma {\rm Pb}}(+y)} +  n_{0{\rm NX}{\rm N}}(\gamma, -y) \cdot {\sigma_{\gamma {\rm Pb}}(-y)}  \text{ ,}\\
		\frac{{\rm d}\sigma^{{\rm XNX}{\rm N}}_{{\rm Pb}{\rm Pb}}}{{\rm d}y}  &= n_{{\rm XNX}{\rm N}}(\gamma, +y) \cdot {\sigma_{\gamma {\rm Pb}}(+y)} +  n_{{\rm N}{\rm XNX}}(\gamma, -y) \cdot {\sigma_{\gamma {\rm Pb}}(-y)} \text{ ,}
	\end{split}
\end{equation}
where 0N0N, 0NXN and XNXN indicate no neutron emission on either sides of the ALICE detector with respect to the interaction point, neutrons only on one side, and neutrons on both sides, respectively. Alternatively, it is possible to study coherent \jpsi in peripheral collisions as well, and simultaneously use peripheral and ultra-peripheral results to extract the photonuclear cross sections as shown in Eq.~\ref{eq:peripheral-photoproduction}~\cite{Contreras:2016pkc}: 
\begin{equation}
	\label{eq:peripheral-photoproduction}
	\begin{split}
		\frac{{\rm d}\sigma_{{\rm Pb}{\rm Pb}}^{\rm P}}{{\rm d}y} &= n_{\rm P}(\gamma, +y)\cdot { \sigma_{\gamma {\rm Pb}}(+y)} + n_{\rm P}(\gamma, -y)\cdot {\sigma_{\gamma {\rm Pb}}(-y)} \text{ ,} \\
		\frac{{\rm d}\sigma_{{\rm Pb}{\rm Pb}}^{\rm U}}{{\rm d}y} &= n_{\rm U}(\gamma, +y)\cdot {\sigma_{\gamma {\rm Pb}}(+y)} + n_{\rm U}(\gamma, -y)\cdot {\sigma_{\gamma {\rm Pb}}(-y)}\text{ ,}
	\end{split}
\end{equation}
where the symbols P and U indicate peripheral and ultra-peripheral samples, respectively.
The results for the extraction of the photonuclear cross sections using the peripheral and neutron emission techniques   are shown in Fig.~\ref{fig:xsec-energy}. The blue empty data points show the results obtained using Run 1 data for both ultra-peripheral and peripheral samples~\cite{Contreras:2016pkc}, while the black data points are the latest ALICE results using Run 2 data in neutron emission classes. The neutron emission results extend the energy dependence coverage by about 300 \GeV, down to a Bjorken-$x$ of about $10^{-5}$. The results obtained with both techniques agree within uncertainties. The most recent results with neutron emission agree with \starlight at low energies, while at higher energies they agree with models including nuclear shadowing and gluon saturation phenomena. Fig.~\ref{fig:shadowing} shows instead the results obtained for the nuclear suppression factor, which is the experimentally accessible observable for the nuclear shadowing factor, and is defined as~\cite{Guzey:2020ntc}:
\begin{equation}
	\label{eq:suppression}
        S_{\rm Pb}(x) = \sqrt{\frac{{\rm d}\sigma_{\rm PbPb}^{\rm data}}{{\rm d}\sigma_{\rm PbPb}^{\rm IA}}} \text{ ,}
 \end{equation}
 where $S_{\rm Pb}(x)$ indicates the suppression factor as a function of Bjorken-$x$, ${\rm d}\sigma_{\rm PbPb}^{\rm data}$ is the measured photonuclear cross section, and ${\rm d}\sigma_{\rm PbPb}^{\rm IA}$ is the prediction from the Impulse Approximation computations. The results obtained by ALICE show a decrease of the suppression factor from a value of about 0.9 at high Bjorken-$x \sim 10^{-2}$ down to about 0.5 at $x \sim 10^{-5}$, in agreement with models including nuclear shadowing and saturation phenomena.

\section{Conclusions}
The ALICE Collaboration has collected a large number of photoproduced \jpsi candidates during LHC Run 2. The collected sample made it possible to have more differential measurements, such as the study of the rapidity dependence of coherent \jpsi production, thus ultimately establishing the presence of significant nuclear shadowing in \PbPb events. It was also possible to measure the $|t|$-dependence of (in)coherent \jpsi production, thus showing how QCD dynamical effects and subnucleonic degrees of freedom are required to describe the data. The ALICE Collaboration also reports its first measurement of coherent \jpsi production in neutron emission classes, which made it possible to study the energy dependence of the photonuclear cross section and the nuclear suppression factor in a wide range from about 20~\GeV up to 800~\GeV, or equivalently from a Bjorken-$x\sim 3\cdot 10^{-2}$ down to $x\sim 10^{-5}$. The results are in agreement at the highest energies with models including nuclear shadowing and saturation phenomena.



\end{document}